\documentclass[conference]{IEEEtran}
\IEEEoverridecommandlockouts
\usepackage{cite}
\usepackage{amsmath,amssymb,amsfonts}
\usepackage{algorithmic}
\usepackage{graphicx}
\usepackage{textcomp}
\usepackage{xcolor}
\def\BibTeX{{\rm B\kern-.05em{\sc i\kern-.025em b}\kern-.08em
    T\kern-.1667em\lower.7ex\hbox{E}\kern-.125emX}}

\author{\IEEEauthorblockN{{Liqiang Yu}$^1$}
\IEEEauthorblockA{\textit{The University of Chicago} \\
Chicago, USA \\
rexyu@uchicago.edu}
\and
\IEEEauthorblockN{Chen Li$^2$}
\IEEEauthorblockA{\textit{The University of Texas at Dallas} \\
Dallas, USA \\
cxl167330@utdallas.edu}
\and
\IEEEauthorblockN{Bo Liu$^3$}
\IEEEauthorblockA{{ZheJiang University}\\
Hangzhou, China \\
21851111@zju.edu.cn}
\and
\IEEEauthorblockN{Chang Che$^*$}
\IEEEauthorblockA{\textit{The George Washington University} \\
Washington, USA \\
cche57@gwmail.gwu.edu}}
\title{Stochastic Analysis of Touch-Tone Frequency Recognition in Two-Way Radio Systems for Dialed Telephone Number Identification}
\begin{document}
\maketitle
\begin{abstract}
This paper focuses on recognizing dialed numbers in a touch-tone telephone system based on the Dual Tone Multi-Frequency (DTMF) signaling technique with analysis of stochastic aspects during the noise and random duration of characters. Each dialed digit's acoustic profile is derived from a composite of two carrier frequencies, distinctly assigned to represent that digit. The identification of each digit is achieved by pinpointing the frequency pair with the highest energy or amplitude in its spectral output, utilizing the Discrete-Time Fourier Transform (DTFT). This analysis includes simulations that illustrate the effects of introducing stochastic variations during the ``Mark" and ``Space" intervals of the decoding process, offering insights into the technique's efficacy and the impact of random temporal fluctuations. This will reduce the accuracy of decoder decoding and lower the SNR of the system.
\end{abstract}


\section{Introduction}

The advent of the Touch-Tone telephone system by AT\&T marked a significant milestone in the evolution of telecommunication technologies, heralding the transition from the traditional rotary dialing system to a more advanced interface. Endorsed as the industry benchmark following the ITU-R recommendation for terrestrial telephony services, this innovative system epitomized a paradigm shift in user interaction with telecommunication system\cite{zhibin2019labeled,zou2023joint}. Central to the Touch-Tone infrastructure is the employment of Dual-Tone Multi-Frequency (DTMF) signaling, a method that facilitates telecommunication signaling across analog lines within the voice-frequency band, bridging telephone handsets with the switching centers and other communication apparatus\cite{zou2017labeled}. In this scheme, each numeral on the telephone keypad is encoded via a pair of tones, derived from a predefined frequency set. These frequency pairs, when transmitted sequentially, are deciphered by the multi-frequency receivers located at the distant telephone exchanges. Initially, the decoding of DTMF tones was achieved through the application of tuned filter banks. Nonetheless, advancements in digital signal processing have led to the predominance of frequency-domain transform techniques, notably the Fast Fourier Transform (FFT), as a more efficient alternative for tone decoding\cite{oppenheim1997signals,proakis2006dimitris}.

The Fourier transform, a cornerstone in signal processing theory, offers a robust framework for converting signals from the time domain to the frequency domain. This frequency domain representation is instrumental in the analysis and interpretation of signals, facilitating the design of filters to suppress unwanted frequencies, examining system responses, and solving differential and difference equation-based system outputs among other applications\cite{10131248,che2023enhancing,gao2023autonomous,mo2022trafficflowgan,shen2023application,wei2024strategic}. However, the operation of such systems is not devoid of stochastic variations. Notably, the digital encoding of frequencies to denote ``Mark" (the tone's duration) and ``Space" (the inter-digit silence) introduces randomness. This variability manifests in the non-uniform length of digits and the inconsistent spacing between them, challenging the assumption of precision in signal transmission. This paper delves into the implications of noise introduction on signal input and examines the resultant effects of stochastic variations in the ``Mark" and ``Space" durations generated by the encoding process. 

\section{Analytical Perspectives on Dual-Tone Multi-Frequency Signaling}

For the four extra digits “A B C D”, we can not see them in the keyboard directly, but they do exit in real. These all use the same row frequencies as a standard keypad, but they have an additional column tone. The extra codes are very useful in preventing standard telephone codes from being used to control remote devices, and can give you override status when used correctly in a two-way radio system.

\begin{table}[htbp]
\caption{DTMF FREQUENCIES FOR TOUCH-TONE TELEPHONES.}
\begin{center}
\begin{tabular}{|c|c|c|c|c|c|}
\hline
\textbf{}&\textbf{1209 Hz}&\textbf{1336 Hz}&\textbf{1477 Hz}&\textbf{1633 Hz}\\
\cline{1-5} 
\textbf{679 Hz} & \textbf{1}& \textbf{2}& \textbf{3}&\textbf{A} \\
\hline
\textbf{770 Hz} & \textbf{4}& \textbf{5}& \textbf{6}&\textbf{B} \\
\hline
\textbf{852 Hz} & \textbf{7}& \textbf{8}& \textbf{9}&\textbf{C}  \\
\hline
\textbf{941 Hz} & \textbf{$*$}& \textbf{0}& \textbf{\#}&\textbf{D}  \\
\hline
\end{tabular}
\label{tab1}
\end{center}
\end{table}

\section{ANALYSIS FOR FREQUENCY}
One of the most potent and popular mathematical tools employed in many areas, i.e., signal analysis, computer vision, statistical analysis, and so on, is the Fourier theory. As a mathematical transform that decomposes functions depending on space or time into functions depending on spatial or temporal frequency, such as the expression of a musical chord in terms of the volumes and frequencies of its constituent notes, it transfers the signal from the time domain to the frequency domain. A periodic signal, whether continuous-time or discrete-time, can be intricately deconstructed into its constituent frequency components. This process enables the utilization of frequency domain representations for critical operations such as signal detection and the selective filtration of desirable signals from those deemed undesirable. The realm of frequency domain analysis, inclusive of spectral estimation, plays a pivotal role across a myriad of scientific and engineering disciplines. These applications span speech processing and recognition, to pivotal medical technologies like Electrocardiography (ECG) and Electroencephalography (EEG), alongside digital filter design, signal denoising, data compression, encryption, radar target identification, and advanced image processing techniques for texture and pattern recognition\cite{wu2024switchtab,gay1989algorithms,mo2022quantifying}. The Fourier transform analysis, in particular, has garnered significant interest and profound scrutiny following the advent of the Fast Fourier Transform (FFT). The FFT algorithm marks a significant breakthrough by drastically reducing the computational demand for an N-point Discrete Fourier Transform (DFT) from a time complexity of $N^2$ operations to a more efficient $N log_2N$ operations, thereby revolutionizing computational efficiency in digital signal processing \cite{zou2022unified,zhang2020manipulator,stylianou2000simple}.

Here we set $x[n]$ as the discrete-time signal, obtained by discretizing the continuous-time variables $x(t)$. The Discrete Time Fourier Transform (DTFT) of the signal $x[n]$ is given by
\begin{equation}
    X(e^{jw}) = \sum_{n = -\infty}^{\infty} x[n] e^{-jwn}
    \label{eq1}
\end{equation}

The Discrete-Time Fourier Transform (DTFT), denoted as $X(e^{j\omega})$, emerges as a frequency-dependent function through its foundational definition, manifesting for signals characterized either by absolute summability or possessing finite energy. The implementation of the DTFT in a numerical context leverages the Fourier Transform, facilitating its computation solely at discrete frequency samples of the continuous variable $\omega$. Given the periodic nature of $X(e^{j\omega})$, an optimal strategy for frequency sampling involves selecting equidistant points within the range $0 \leq \omega \leq 2\pi$, defined by $ \omega_c = \frac{2\pi k}{N} $, where $ k = 0,1,...,N - 1 $. In the scenario of a finite duration signal $x[n]$ with length $M$, and non-zero only for the interval $0 \leq n \leq M - 1$, these frequency samples yield the transformation as
\begin{equation}
    X(e^{jw_k}) = \sum_{n = 0}^{M - 1} x[n] e^{-j2\pi kn/N}
    \label{eq2}
\end{equation}

\section{GENERATION OF DTMF DIGIT}\label{AA}
In the context of Dual-Tone Multi-Frequency (DTMF) signaling within a touch-tone telephone system, the actuation of any key instigates the generation of two distinct tones. These tones, comprising sinusoids of higher and lower frequencies, are systematically assigned to the matrix of keys on the touchpad, with higher frequencies allocated to columns and lower frequencies to rows. Illustratively, Table II delineates the configuration of a telephone keypad, mapping each digit to a pair of DTFT frequencies. This mapping assumes the dual-tone waveform is discretely sampled at a frequency of $F_s = \dfrac{1}{T_s} = 8192 Hz$, a sampling rate predominantly employed in the conveyance of speech or voice signals across telephonic networks\cite{zhao2023stable}. As a case point, consider the digit '2', which is represented by two discrete frequencies: $f_{\text{row}} = 697 Hz$ and $f_{\text{column}} = 1336 Hz$. Consequently, the composite tone signal for this digit is articulated as
\begin{equation}
    \begin{aligned}
        d_2(t) &= sin(2 \pi f_\text{row}t) + sin(2\pi f_\text{column}t)\\
        &= sin(2 \pi (697)t) + sin(2 \pi (1336)t)
    \end{aligned}
    \label{eq3}
\end{equation}
After the discretization process as $t = nT_s = n/F_S$, we can get the discrete time signal for $digit 2$ is
\begin{equation}
    d_2[n] = sin(0.534n) + sin(1.0247n)
    \label{eq4}
\end{equation}
Furthermore, it is feasible to ascertain the discrete-time signals correlated with additional numerical digits. Table II delineates the frequencies, both low and high, expressed in radians, pertinent to these digits.
\begin{table}[htbp]
\caption{DTFT FREQUENCIES FOR TOUCH TONE SIGNALS SAMPLED AT
FREQUENCY 8192 Hz}
\begin{center}
\setlength{\tabcolsep}{6mm}{}
\begin{tabular}{|c|c|c|c|}
\hline
\textbf{}&\multicolumn{3}{|c|}{$\boldsymbol{\omega}_{\textbf{column}}$} \\
\cline{1-4} 
$\boldsymbol{\omega}_\textbf{row}$ & \textbf{0.927}& \textbf{1.0247}& \textbf{1.1328} \\
\hline
\textbf{0.5346} & \textbf{1}& \textbf{2}& \textbf{3} \\
\hline
\textbf{0.5906} & \textbf{4}& \textbf{5}& \textbf{6} \\
\hline
\textbf{0.6535} & \textbf{7}& \textbf{8}& \textbf{9} \\
\hline
\textbf{0.7217} & \textbf{}& \textbf{0}& \textbf{} \\
\hline
\end{tabular}
\label{tab2}
\end{center}
\end{table}

\section{STOCHASTIC ASPECTS ANALYSIS}
\subsection{CONCEPTS FOR MARK AND SPACE}
Before the talk in this section, we need to know two important concepts for the encoder: "Mark" and "SPACE". Mark and Space refer to the duration a DTMF tone is produced, as well as the duration of the silence between individual digits. The time which a DTMF digit tone is actually producing sound, is called the “Mark” time. The silence between each one of the digits is called the “Space”. Most DTMF decoders and controllers will list a minimum Mark/Space speed, expressed in milliseconds. Mark/Space is pronounced “Mark and Space” not “Mark divided by Space”. It is not a ratio between the two speeds, but rather the duration of each step in the DTMF code. The standard for most radio decoders and most telephone equipment is 40/40. The decoders expect the DTMF tones to existing for at least 40 milliseconds, with 40 milliseconds of silence between each DTMF digit. While it is generally accepted to have the Space duration be the same as the Mark, this is unnecessary. The purpose of the Space is to give the decoder notice that a DTMF digit just ended. A Space need only be long enough for it to accomplish that purpose. A Space can be 5 msec or less on a fast decoder and still perform its function. While a 40/40 timing is just fine, if max speed is necessary, you can consider adjusting the Space timing of your encoder to a faster rate. Here we can see the the "Fig. 1" which can be specifically shown the information of Mark and Space.  

\begin{figure}[htbp]
\centerline{\includegraphics[height=3cm]{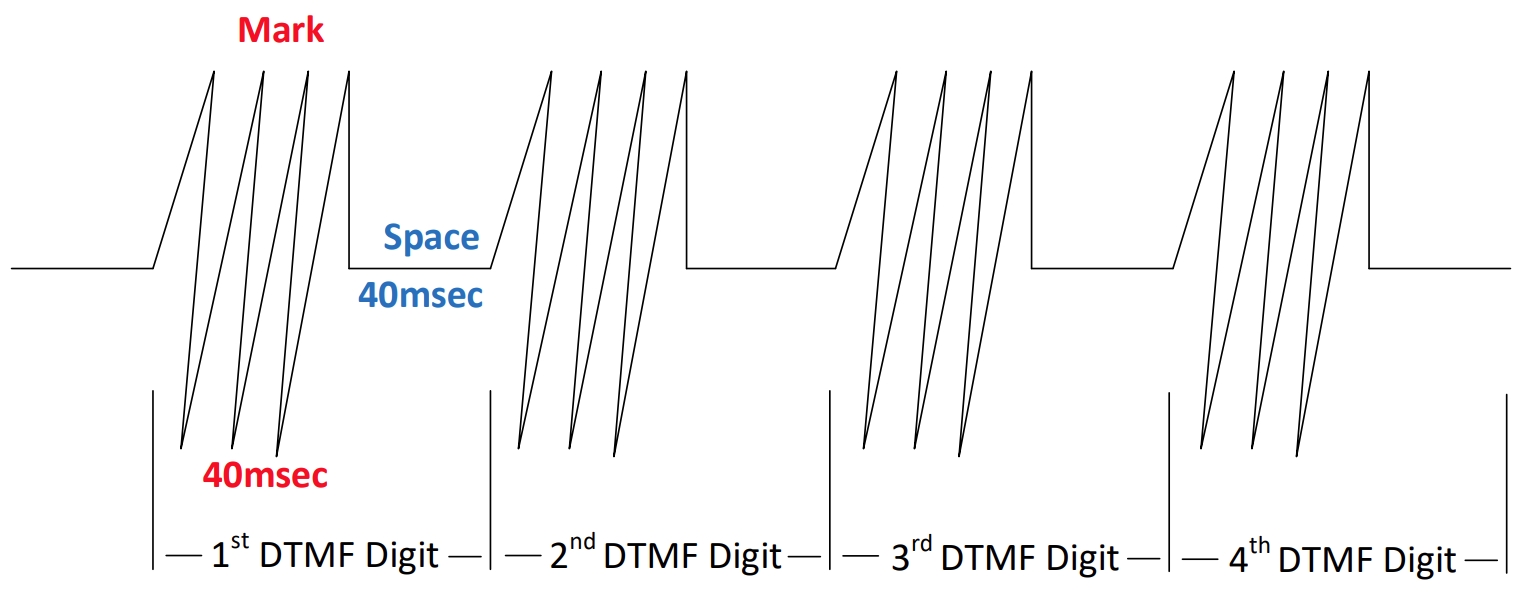}}
\caption{Explanation of DTMF timing with ``Mark" and ``Space" in the ideal situation.}
\label{fig}
\end{figure}

However, this situation is only for ideal design or condition. In reality, we can not get specified timing for Mark and Space with ruthless precision. Each digit doesn't have the same length, and the space between digits is always the same length. So that means the stochastic aspects such as noise and random duration of characters should also be considered. The code generated by encoder with adding noise on each digit will influence the rightness of the decoder. For DTFT method, the number of samples for each digit will also influence the rightness of the decoder. We can show the effect as the noise and stochastic added on the encoder in the next two subsections.

\subsection{STOCHASTIC ASPECTS IN AUTOMATIC DTMF ENCODERS}
To send the digits quickly, the operator need to be using an automated encoder since we could not press each button by hand that fast. An automated encoder is also called a ``Store and Forward” encoder. You ``Store” the DTMF code into the device when you program it, then ``Forward” it to send the entire DTMF code as one complete DTMF string of numbers and characters. Store and forward encoders can be a programmable microphone or two-way radio, a stand-alone encoder such as the ENC-3300, or part of a large law enforcement dispatch console. In the end they all perform the same function of sending out DTMF digits and characters.\\

For encoder, when the code generated with different timing Mark with random changes, that means that each digit lasts for a different amount of time. This is undoubtedly a challenge for the correctness of the decoder because most of the decoders among the companies do not have extraordinarily stable and fault-tolerant algorithms to compensate for this problem. Also the signal-to-noise ratio(SNR), which can be represent as
\begin{equation}
    SNR = 10log_{10} \dfrac{\sigma{s(t)^2}}{\sigma_{h(t)-s(t)}^2}
    \label{eq5}
\end{equation}
where $\sigma_{h(t)-s(t)}^2$ denotes the variance of the modeling error, and the $\sigma_{s(t)}^2$ denotes the variance of the original speech signal $s(t)$. For the output of the encoder would be influenced that can also decrease the rightness of the decoder. we will show the experiment about the effect of different noise added on specified digit in the next chapter.   
\subsection{STOCHASTIC ASPECTS IN MANUAL DTMF ENCODERS}
We can explore from the following directions the influence of the stochastic aspect on the touch-tone telephone system.
\subsubsection{The original input analog signal on an encoder is disturbed by white noise}
For the original input analog signal on the encoder, we can use the Eq. (6) to represent any digits from "0-9". 
\begin{equation}
    d_n[n] = sin(w_\text{row} \cdot n) + sin(w_\text{column}\cdot n)
    \label{eq6}
\end{equation}

Based on the Eq. (6), we can add Gaussian white noise on the signal and see the influence on the each digit signal. We would talk about this part in detail during the VII section. 
\subsubsection{Change the preset duration of Mark and Space randomly when the decoder decodes the phone numbers generated and stored by the encoder} For automatic DTMF encoder, it require the encoder that you can store the DTMF code into the device when you program it, then “Forward” it to send the entire DTMF code as one complete DTMF string of numbers and characters. In this case, the information generated and stored by the encoder is fixed, and the mission for the decoder is to analyze the data and decode them to the correct digit signal that a touch-tone telephone system can use to do the subsequent operations\cite{ni2024smartfix}. In this section we can change the length of the interval used by the decoder to parse 
Mark and Space timing during the information generated and stored by the encoder. 
\subsubsection{Randomly change the sampling interval of each digit in the frequency domain when the encoder performs a DTFT transformation of the original input analog signal}
When we use the DTFT method to compute N samples of the DTFT of a finite-length signal as frequency $w_k = 2 \pi k/N$,It intuitively converts the superimposed sine wave signal into a double-peaked signal. The decoder can decode the information generated by the decoder by comparing the two DTFT frequencies corresponding to the peak with preset data based on Table II. \\\\
We will talk about this part in the next chapter and show related experiments to explain the each part in detail.

\section{THE DECODING ALGORITHM OF DTMF}
Due to decoder cannot identify the DTMF signals by sampling time signals directly, we can analyze and determine through frequency domain. Here we mainly introduce two decoding algorithms and compare them during the experiment with stochastic aspect adding on the encoder in the next chapter. 
\subsection{The Goertzel algorithm}
The Goertzel algorithm as an effective convolutional form of the discrete Fourier transform for direct computation of the Fourier value at selected frequency position, i.e., evaluates the only selected bin of the Fourier spectrum. The basic relation of the discrete Fourier transform Eq. (2) can be also represent as
\begin{equation}
    X(k) = \sum_{r = 0}^{N - 1} x(r) e^{-j2\pi kr/N}
    \label{eq7}
\end{equation}

Then we put the Eq. (7) in to a convolutional Eq. (8), where $x(n)$ is an $n^{th}$ signal sample in time domain and $X(k)$ is a $k^{th}$ bin of the Fourier spectrum.
\begin{equation}
    \begin{aligned}
    y_k(n) &= \sum_{r = -\infty}^{\infty} x(r) \cdot e^{j2\pi(N - r)\dfrac{k}{N}} \cdot u(n - r) \\
           &= x(n) * e^{j2\pi nk/N}|_{n = N}
\end{aligned}
\label{eq8}
\end{equation}

An impulse response of the derived filter is then a complex harmonic signal Eq. (9) of which length is constrained by a rectangular window.
\begin{equation}
    h(n) = e^{j2\pi n k/N}
\end{equation}

By applying the Z-transform to the impulse response (7), it is possible to find the transfer function of the Goertzel filter(8). More convenient for implementation of the Goertzel algorithm is the modified form (9), which can be split into the real recursive and the complex direct computational parts.
\begin{equation}
\begin{aligned}
    H(z) &= \sum_{n = 0}^{\infty} h(n)z^{-1} = \dfrac{1}{1 - z^{-1}e^{j2 \pi k/N}}\\
         &= \dfrac{1 - z^{-1} e^{-j2 \pi k/N}}{1 - 2z^{-1}cos(2 \pi k/N) + z^{-2}}
\end{aligned}
\label{eq9}
\end{equation}
We show the realization of the transfer function (8) in “Fig. 2”. Notice that the filter has two complex poles located on the unit circle that is a condition of stability. Indeed, the loop can generate an impulse response with stable amplitude. Some uncertainty of the coefficient C caused, e.g., by a quantization error, may fade the impulse response, which can be noticeable, especially in the case of significant coefficients both k and N. The down-sampling blocks stand for the condition in  \cite{zhao2024nlbac} and provide the rectangular windowing of the impulse response. The direct part of the algorithm evaluates the complex output. Here we can show the realization of the transfer function Eq. 8 in Fig. 2 as the structure of the Goertzel algorithm. 
\begin{figure}[htbp]
\centerline{\includegraphics[height=4cm]{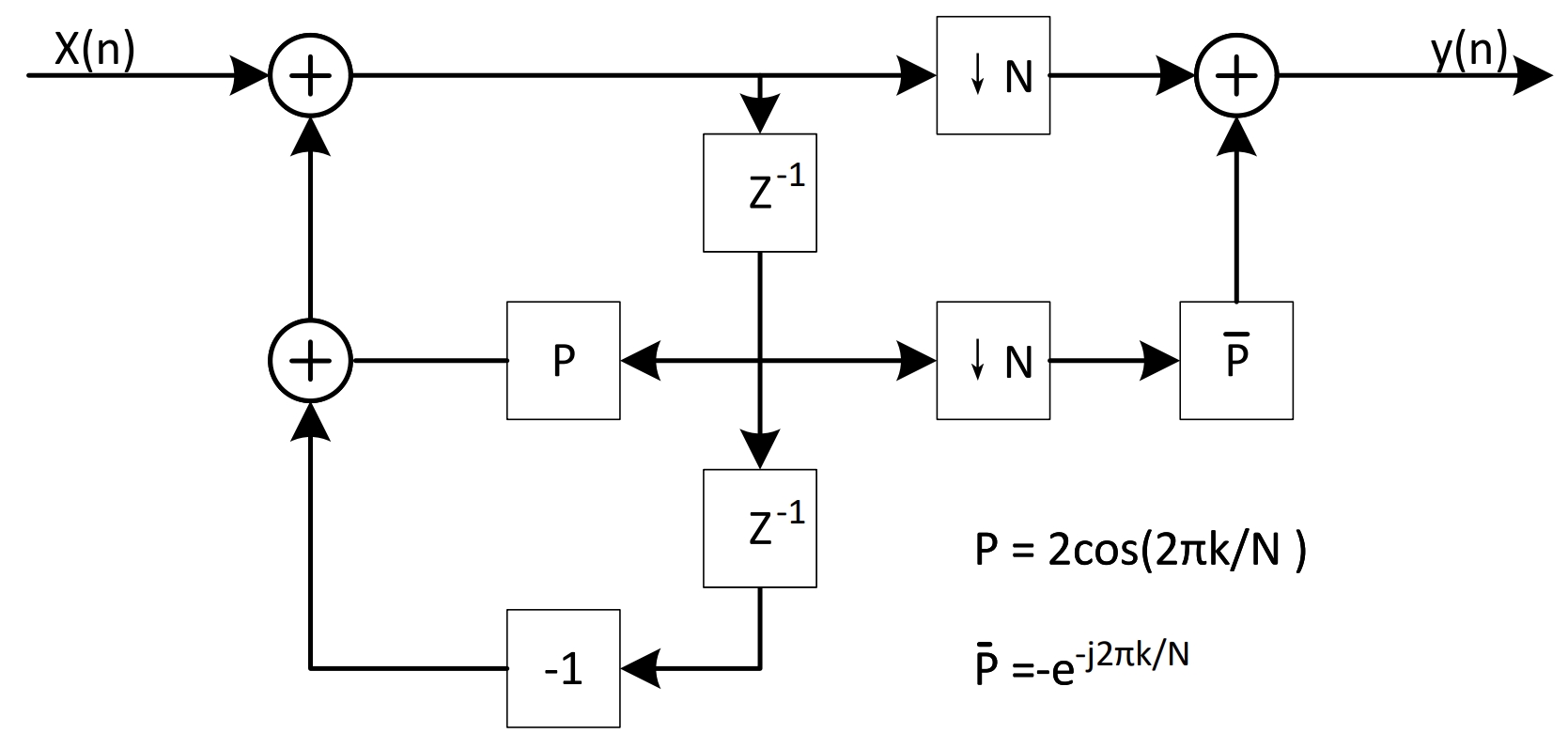}}
\caption{The structure of the Goertzel filter.}
\label{fig}
\end{figure}

\subsection{Algorithm of improved Goertzel based on non-normalized discrete Fourier transform}
The non-normalized discrete Fourier transform define the sequence x[n] which the length is N as

$$ X(z_k) =  \sum_{n = 0}^{N - 1} x(n) z_k^{-n},\qquad k = 0,1,...,N - 1 \eqno{(11)}$$

where $z_0, z_1,...,z_{N-1}$ are distinct points locacated arbitrarily in the z-plane. This is also equivalent to the second approach described in  \cite{chen2023recontab}, where the modified Goertzel algorithm is used for implementation.

To compute NDTF, we need to use sub-band NTDF(SB-NDFT) based on a sub-band decomposition of the input sequence. It is useful for developing fast algorithms to compute NDFT samples approximately, for signals which have their energy concentrated in a few bands of the spectrum. Consider a sequence z[n] with an even number of samples N. We first decompose z[n] into two sub-sequences, g~[n] and g~[n], of length N/2 each:
\begin{align*}
   g_L(n) &= \dfrac{1}{2}\{ x[2n] + x[2n + 1] \}, \tag{12} \\
   g_H(n) &= \dfrac{1}{2}\{ x[2n] - x[2n + 1] \}, \tag{13}
\end{align*}

We can organize the form to express x[n] in terms of $g_L[n]$ and $g_H[n]$ as
\begin{align*}
   x[2n] &= g_L[n] + g_H[n], \tag{14} \\
   x[2n + 1] &= g_L[n] - g_H[n], \tag{15}
\end{align*}

By substituting for z[n] in the NDFT expression in Eq. (11), we can obtain 
\begin{align*}
    X(z_k) &= \sum_{n = 0}^{N/2 - 1} x[2n]z_k^{-2n} + \sum_{n = 0}^{N/2 - 1} x[2n + 1]z_k^{-(2n + 1)} \\
           &= (1 + z_k^{-1})G_L(z_k^2) + (1 - z_k^{-1})G_H(z_k^2),  \tag{16}
\end{align*}
where $G_L(z_k^2)$ and $G_H(z_k^2)$ are the NDFTs of the sequences, $g_l[n]$ and $g_H[n]$, evaluated at $z = z_k^2$. Therefore, we can find the NDFT of x[n] at the point $z = z_k$ by computing the NDFTs of the samller sequences, $g_L[n]$ and $g_H[n]$, at the point $z = z_k^2$, and then combining them as shown in Eq. (14). The sampling rate frequency used for telephone is fs=8 kHz, all the DTMF frequencies are located in the low frequency sections of the sampling frequencies $(0 \leq f \leq f_s/4)$.So, the NDFT can be get approximately by reducing the high response sections:
$$ \hat X(z_k) = (1 + z_k^{-1})G_L(z_k^2) \eqno{(17)} $$
The deviation which caused by removing the high frequency part can be compensated as
$$ X(z_k) = \dfrac{2 \hat X(z_k)}{1 + cos w_k} \eqno{(18)} $$
For DTMF decoding, we need only the squared magnitudes of the approximate NDFT samples.
From Eq. (17) and Eq. (18), we obtain
$$ |X(z_k)|^2 = \dfrac{4|\hat X(z_k)|^2}{(1 + cos(w_k))^2} = \dfrac{8|G_L(z_k^2)|^2}{1 + cos(w_k)} \eqno{(19)}$$
The value of $|G_L(z_k^2)|^2$ is computed by using the modified Goertzel algorithm.
\section{SIMULATIONS}
We divides this section into three parts: experiments without stochastic analysis and experiments with stochastic analysis, digital filters used for DTMF decoding. The experiment without stochastic analysis will make the following studies. In the first part, we generate a discrete-time signal containing a specified phone number defined by ourselves with each digit constructed as given in Eq. (3). In the second part, We add stochastic interference to the signal generated in the first part and explains the comparison between the signal after adding stochastic interference and the original signal to determine whether the interference would cause damage to the information of the original signal.In the third part, we will introduce the Goertzel algorithm we mentioned in the last chapter and show the effect of using such a filter method on a specified signal.
\subsection{Experiments without stochastic analysis for DTMF encoder}
For the purpose of digital signal analysis, each numeral within a telephonic sequence is represented through the generation of 1,000 discrete samples. Intervening periods of silence, delineating the separation between successive numerals, are denoted by a sequence of 100 null samples, signifying quiescence. These quantities, 1,000 and 100, are predetermined constants, integral to the experimental setup, and remain invariant throughout the procedure. Subsequently, the composite signal, embodying the telephonic number, is audibly rendered in MATLAB at a predefined sampling frequency of 8,192 samples per second. This auditory output replicates the acoustic experience encountered during the manual input of a phone number via a conventional telephone keypad. The spectral signature of each digit, serving as a benchmark for analysis, is elucidated through the application of the Discrete-Time Fourier Transform (DTFT), as delineated in Equation (3) and further expounded upon in Table II. By way of illustration, Figure 3 delineates the normalized Fast Fourier Transform (FFT) magnitude spectrum for the tonal components of the digit '2', revealing distinct spectral peaks at frequencies $w_{row} = 0.5346$ and $w_{column} = 1.0247$ radians. This a priori knowledge concerning the spectral peak coordinates for each numeral facilitates the decoding process of dialed sequences. Such decoding is achieved by conducting an FFT-based comparative analysis, aligning the actual tonal outputs against the comprehensive spectrum of potential digits as cataloged in Table II.
\begin{figure}[htbp]
\centerline{\includegraphics[height=4cm]{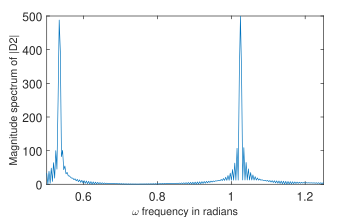}}
\centerline{(a)}
\centerline{\includegraphics[height=3.3cm]{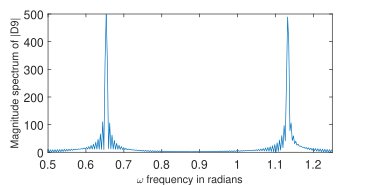}}
\centerline{(b)}
\caption{The magnitude spectrum of (a) $D_2(e^{jwk})$ and (b) $D_9(e^{jwk})$}
\label{fig}
\end{figure}

Figure 4 depicts the amplitude of the digit 2's tone signal in the time domain. By analyzing the spectral peaks of tone signals for each digit, referenced against frequencies in Table II, we can identify dialed digits. This process allows for the accurate decoding of a ten-digit phone number from its spectral data. In the experiment detailed, the decoded number is 247-448-1221, showcasing the method's precision in identifying dialed sequences through spectral analysis.

\begin{figure}[htbp]
\centerline{\includegraphics[height=4cm]{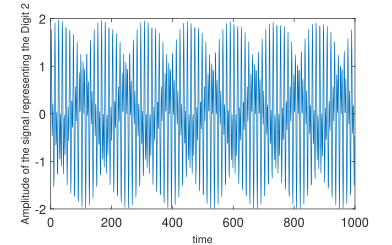}} 
\centerline{(a)}
\centerline{\includegraphics[height=4cm]{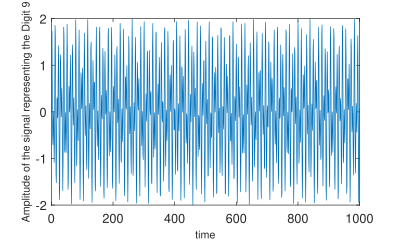}} 
\centerline{(b)}
\caption{Time domain signal of the dialed digit of (a) $D_2(e^{j\omega k})$ and (b) $D_9(e^{j\omega k})$}
\label{fig}
\end{figure}

\subsection{Experiments with stochastic analysis for DTMF encoder}
For stochastic analysis of encoder, we can focus on the following two aspects: (1). White noise corrupts the touch tone signal. (2). Random change for Mark and Space time.
\subsubsection{The original input analog signal on an encoder is
disturbed by white noise}
Consider the scenario where $x[n]$ represents the pristine tone sequence generated upon the actuation of a digit, and $w[n]$ denotes the white noise interference. The resultant tone, compromised by noise, is thus expressed as
$$ y[n] = x[n] + w[n] \eqno{(20)}$$
In this context, let the power of the unblemished signal be denoted by $\sigma_x^2$, and the power of the additive white noise by $\sigma_n^2$. Consequently, the ratio of signal power to noise power (SNR), quantified in decibels (dB), is articulated as
$$ SNR = 20\log_{10}\left(\frac{\sigma_x}{\sigma_n}\right) \eqno{(21)}$$
Here we choose SNR range from 1 to 5 to add the noise on the primary signal. First, we will compare different signal amplitudes representing digit 2 in the time domain with and without noise during SNR from 1 to 5 in Fig. 5. Based on Fig. 5 we can obtain that when then SNR improved, the signal with noise is closer to the original signal. Then, we will show different magnitude spectrum of Digit 2 with SNR from 1 to 5 in Fig. 6. Comparing results with the original signal, we can obtain that the value of the signal-to-noise ratio increases the closer the decoder output signal is to the original signal. The rate of error for each digit when the SNR change is given by
$$ E_{rror} = \dfrac{\sigma_n^2}{\sigma_x^2}\cdot 100\% \eqno{(22)} $$
We can obtain the changing of error for each digit when the SNR was changed in Fig. 7. Based on the graph we got, we can see that each digit has the same error trends by SNR change.  


\begin{figure}[htbp]
\centerline{\includegraphics{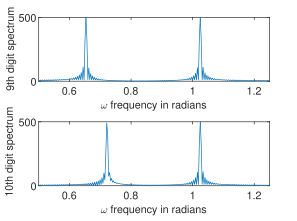}}
\centerline{{(c)}}
\caption{Amplitude spectrum of the dialed digits for the (a) $1^{st}, 2^{nd}, 3^{rd}$ and $4^{th}$ digits (b) $5^{th}, 6^{th}, 7^{th}$ and $8^{th}$ (c) $9^{th}$ and $10^{th}$ digits starting from the top}
\label{fig}
\end{figure}

\begin{figure}[htbp]
\centerline{\includegraphics[height=13cm]{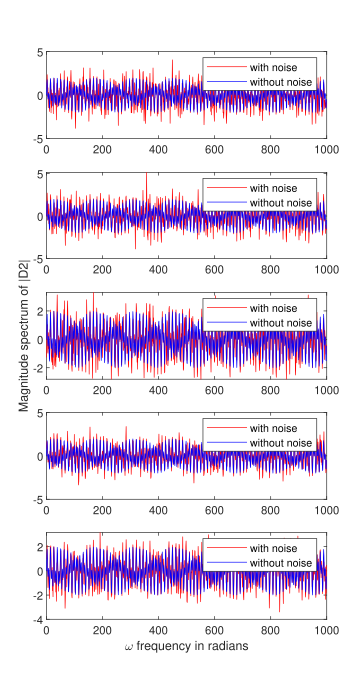}}
\caption{Amplitude of the signal representing digit 2 in time domain with SNR from 1 to 5.}
\label{fig}
\end{figure}


\begin{figure}[htbp]
\centerline{\includegraphics[height=12cm]{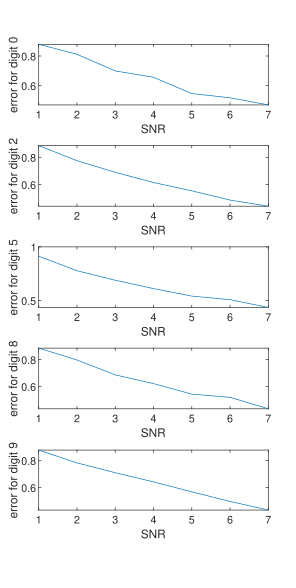}}
\caption{The change of error for the dialed digits 0, 2, 5, 8, 9 with the range of SNR from 1 to 5.}
\label{fig}
\end{figure}
\subsubsection{Change the preset duration of Mark and Space randomly
when the decoder decodes the phone numbers generated and
stored by the encoder}In this part, we change the preset duration of Mark and Space for each digit on the decoder. We set the Mark timing values as 900, 1050, 980, 1300, 680, 900 and 620 denote digits 4, 9, 1, 5, 8, respectively. We also set the Space timing values as 150, 201,21,400,320 and 80 for each interval timing between the two neighbor digits. Here we can see the comparison of the digit spectrum between the fixed duration and changed duration in Fig. 8. Based on the result we can obtain that even though the Mark and Space timing change synchronously, the distributions of the spectrum peaks in the two situations are nearly the same, only slightly moving from the set DTFT frequency value in TABLE II.

\begin{figure}[htbp]
\centerline{\includegraphics{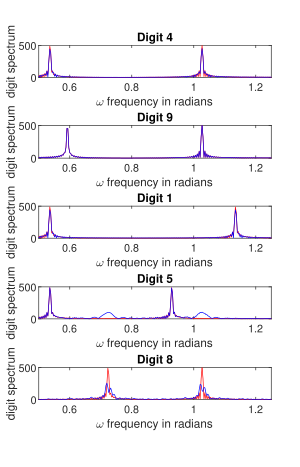}}
\caption{The change of error for the dialed digits 0, 2, 5, 8, 9 with the range of SNR from 1 to 5.}
\label{fig}
\end{figure}

\subsubsection{Randomly change the sampling interval of each digit in the frequency domain when the encoder performs a DTFT
transformation of the original input analog signal}
In the previous experiments, we set both the duration of Mark and Space as stable variables with 100 and 1000, respectively, without any changes. When we use the DTFT method to analyze samples of the original finite-length signal, the N number is fixed as 2048 without any changes. In this experiment, we would change Mark and Space time as two random distributions with the mean value of 2048 and 100, respectively.

We choose seven values for different N variable between 1000 and 3000 as sampled variables when the encoder uses the DTFT method to compute N samples of the DTFT of a finite-length signal at frequencies $w_k = 2 \pi k /N$. In matlab, we generate the sequence of N as: 1100, 1300, 1500, 1700, 2500, 2800, 1900. As with the methods and operations in previous experiments, we can obtain a series of finite-length signals: digit ``4", digit ``9", digit ``1", digit ``5" and digit ``8" with different N samples in different frequencies ranges. Here we can show the new magnitude spectrum of each digit in Fig. 8. Then we compare the magnitude spectrums of each digit in fixed sampled ranges and different sampled ranges based on the random distribution.

During the comparison, we can obtain that when the sampled ranges are changed, the peaks in these two situations fall at the frequencies specified in TABLE II with little difference while keeping full vertical range. Consequently, we can obtain that the change of the sample ranges cannot damage the primary information of each digit. 


\subsubsection{digital filter used for DTMF decoding}
Spectral analysis reveals that the decoding of Dual-Tone Multi-Frequency (DTMF) tones, associated with inputted digits, is achievable through the identification of peak amplitudes within the DTMF signal's frequency spectrum. The design of a second-order Infinite Impulse Response (IIR) filter, characterized by the strategic placement of its poles proximate to the target DTMF frequencies to enhance gain, alongside zeros positioned to attenuate or exclude undesired frequencies, facilitates this process. This approach underscores the utility of the Goertzel algorithm for the extraction of Discrete Fourier Transform (DFT) components from a signal $x(n)$ via a series of parallel filters. The methodology is notably efficient when the objective is to ascertain specific frequency components rather than the signal's complete fundamental frequency spectrum.

Illustrative of this technique is the frequency response of eight bandpass filters, each aligned with the DTMF standard frequencies (697 Hz, 770 Hz, 852 Hz, 941 Hz, 1209 Hz, 1336 Hz, 1477 Hz, and 1633 Hz). The proximity of the response peaks to the respective DTMF frequencies validates the efficacy of this filtering approach in tone detection.

\section*{CONCLUSION}
Within the scope of this manuscript, we delve into the analytical approach for decoding dialed numbers within the framework of a touch-tone telephone system, leveraging the frequency spectrum corresponding to the tones of each digit. The Dual-Tone Multi-Frequency (DTMF) signaling protocol serves as the foundation for digit representation, with distinct frequencies attributed to each numeral. Empirical findings detailed in section VII underscore the efficacy of the Discrete-Time Fourier Transform/Fast Fourier Transform (DTFT/FFT) technique in digit decoding, facilitated through a comparative analysis with a reference frequency spectrum for each digit. Furthermore, the manuscript elucidates the decoding algorithm for a predefined telephone number via MATLAB simulations, incorporating a probabilistic evaluation of the encoding process by examining the effects of white noise and digit duration variability on the decoding accuracy. This exploration extends to the assessment of system performance under varied durations of Space and Mark, illustrating the comprehensive decoding capability of the DTFT/FFT approach, even in the presence of stochastic perturbations. The possibility of employing Digital Signal Processing (DSP) algorithms or the design of specialized digital and analog bandpass filters for DTMF tone decoding is also addressed. Given the burgeoning reliance on DTMF-based remote control systems amidst the rapid proliferation of telecommunications, particularly within mobile networks, the imperative for flawless DTMF decoding becomes paramount. Prospective research endeavors will focus on harnessing the DTMF protocol for enhanced control applications\cite{chen2023recontab,chen2018data,mo2022uncertainty,chen2017personalized,zang2024precision}.


\bibliographystyle{ieeetr}
\bibliography{main}

\begin{thebibliography}{10}

\bibitem{zhibin2019labeled}
Z.~Zhibin, S.~Liping, and C.~Xuan, ``Labeled box-particle cphd filter for multiple extended targets tracking,'' {\em Journal of Systems Engineering and Electronics}, vol.~30, no.~1, pp.~57--67, 2019.

\bibitem{zou2023joint}
Z.~Zou, M.~Careem, A.~Dutta, and N.~Thawdar, ``Joint spatio-temporal precoding for practical non-stationary wireless channels,'' {\em IEEE Transactions on Communications}, vol.~71, no.~4, pp.~2396--2409, 2023.

\bibitem{zou2017labeled}
Z.-b. Zou, L.-p. Song, and Z.-l. Song, ``Labeled box-particle phd filter for multi-target tracking,'' in {\em 2017 3rd IEEE International Conference on Computer and Communications (ICCC)}, pp.~1725--1730, IEEE, 2017.

\bibitem{oppenheim1997signals}
A.~V. Oppenheim, A.~S. Willsky, S.~H. Nawab, and J.-J. Ding, {\em Signals and systems}, vol.~2.
\newblock Prentice hall Upper Saddle River, NJ, 1997.

\bibitem{proakis2006dimitris}
J.~G. Proakis, ``Dimitris. g, manolakis,'' {\em Digital Signal Processing, Prentice Hall Inc}, 2006.

\bibitem{10131248}
H.~Zou, R.~Yu, R.~Anand, J.~Tong, and A.~Q. Huang, ``A gan variable-frequency series resonant dual-active-bridge bidirectional ac-dc converter for battery energy storage system,'' in {\em 2023 IEEE Applied Power Electronics Conference and Exposition (APEC)}, pp.~150--157, 2023.

\bibitem{che2023enhancing}
C.~Che, Q.~Lin, X.~Zhao, J.~Huang, and L.~Yu, ``Enhancing multimodal understanding with clip-based image-to-text transformation,'' in {\em Proceedings of the 2023 6th International Conference on Big Data Technologies}, pp.~414--418, 2023.

\bibitem{gao2023autonomous}
L.~Gao, G.~Cordova, C.~Danielson, and R.~Fierro, ``Autonomous multi-robot servicing for spacecraft operation extension,'' in {\em 2023 IEEE/RSJ International Conference on Intelligent Robots and Systems (IROS)}, pp.~10729--10735, IEEE, 2023.

\bibitem{mo2022trafficflowgan}
Z.~Mo, Y.~Fu, D.~Xu, and X.~Di, ``Trafficflowgan: Physics-informed flow based generative adversarial network for uncertainty quantification,'' in {\em Joint European Conference on Machine Learning and Knowledge Discovery in Databases}, pp.~323--339, Springer, 2022.

\bibitem{shen2023application}
Z.~Shen, K.~Wei, H.~Zang, L.~Li, and G.~Wang, ``The application of artificial intelligence to the bayesian model algorithm for combining genome data,'' {\em Academic Journal of Science and Technology}, vol.~8, no.~3, pp.~132--135, 2023.

\bibitem{wei2024strategic}
K.~Wei, H.~Zang, Y.~Pan, G.~Wang, and Z.~Shen, ``Strategic application of ai intelligent algorithm in network threat detection and defense,'' {\em Journal of Theory and Practice of Engineering Science}, vol.~4, no.~01, pp.~49--57, 2024.

\bibitem{wu2024switchtab}
J.~Wu, S.~Chen, Q.~Zhao, R.~Sergazinov, C.~Li, S.~Liu, C.~Zhao, T.~Xie, H.~Guo, C.~Ji, {\em et~al.}, ``Switchtab: Switched autoencoders are effective tabular learners,'' {\em arXiv preprint arXiv:2401.02013}, 2024.

\bibitem{gay1989algorithms}
S.~L. Gay, J.~Hartung, and G.~L. Smith, ``Algorithms for multi-channel dtmf detection for the we dsp32 family,'' in {\em International Conference on Acoustics, Speech, and Signal Processing,}, pp.~1134--1137, IEEE, 1989.

\bibitem{mo2022quantifying}
Z.~Mo, Y.~Fu, and X.~Di, ``Quantifying uncertainty in traffic state estimation using generative adversarial networks,'' in {\em 2022 IEEE 25th International Conference on Intelligent Transportation Systems (ITSC)}, pp.~2769--2774, IEEE, 2022.

\bibitem{zou2022unified}
Z.~Zou, M.~Careem, A.~Dutta, and N.~Thawdar, ``Unified characterization and precoding for non-stationary channels,'' in {\em ICC 2022-IEEE International Conference on Communications}, pp.~5140--5146, IEEE, 2022.

\bibitem{zhang2020manipulator}
Y.~Zhang, X.~Wang, L.~Gao, and Z.~Liu, ``Manipulator control system based on machine vision,'' in {\em International Conference on Applications and Techniques in Cyber Intelligence ATCI 2019: Applications and Techniques in Cyber Intelligence 7}, pp.~906--916, Springer, 2020.

\bibitem{stylianou2000simple}
Y.~Stylianou, ``A simple and fast way of generating a harmonic signal,'' {\em IEEE Signal Processing Letters}, vol.~7, no.~5, pp.~111--113, 2000.

\bibitem{zhao2023stable}
L.~Zhao, K.~Gatsis, and A.~Papachristodoulou, ``Stable and safe reinforcement learning via a barrier-lyapunov actor-critic approach,'' in {\em 2023 62nd IEEE Conference on Decision and Control (CDC)}, pp.~1320--1325, IEEE, 2023.

\bibitem{ni2024smartfix}
F.~Ni, H.~Zang, and Y.~Qiao, ``Smartfix: Leveraging machine learning for proactive equipment maintenance in industry 4.0,'' in {\em The 2nd International scientific and practical conference “Innovations in education: prospects and challenges of today”(January 16-19, 2024) Sofia, Bulgaria. International Science Group. 2024. 389 p.}, p.~313, 2024.

\bibitem{zhao2024nlbac}
L.~Zhao, K.~Miao, K.~Gatsis, and A.~Papachristodoulou, ``Nlbac: A neural ordinary differential equations-based framework for stable and safe reinforcement learning,'' {\em arXiv preprint arXiv:2401.13148}, 2024.

\bibitem{chen2023recontab}
S.~Chen, J.~Wu, N.~Hovakimyan, and H.~Yao, ``Recontab: Regularized contrastive representation learning for tabular data,'' {\em arXiv preprint arXiv:2310.18541}, 2023.

\bibitem{chen2018data}
S.~Chen, L.~Lu, Y.~Xiang, Q.~Lu, and M.~Li, ``A data heterogeneity modeling and quantification approach for field pre-assessment of chloride-induced corrosion in aging infrastructures,'' {\em Reliability Engineering \& System Safety}, vol.~171, pp.~123--135, 2018.

\bibitem{mo2022uncertainty}
Z.~Mo and X.~Di, ``Uncertainty quantification of car-following behaviors: physics-informed generative adversarial networks,'' in {\em the 28th ACM SIGKDD in conjunction with the 11th International Workshop on Urban Computing (UrbComp2022)}, 2022.

\bibitem{chen2017personalized}
S.~Chen, W.~D. Kearns, J.~L. Fozard, and M.~Li, ``Personalized fall risk assessment for long-term care services improvement,'' in {\em 2017 Annual Reliability and Maintainability Symposium (RAMS)}, pp.~1--7, IEEE, 2017.

\bibitem{zang2024precision}
H.~Zang, ``Precision calibration of industrial 3d scanners: An ai-enhanced approach for improved measurement accuracy,'' {\em Global Academic Frontiers}, vol.~2, no.~1, pp.~27--37, 2024.

\end{thebibliography}

\end{document}